\documentclass[twocolumn]{aa}
\usepackage{graphicx}
\usepackage{txfonts}
\begin{document}
\title{A low-mass stellar companion of the planet host star HD\,75289\thanks{Based on observations obtained
on La Silla in ESO programs 68.C-008 and 70.C-0116(A)}}

\author{M. Mugrauer \inst{1} \and R. Neuh\"auser \inst{1} \and
T. Mazeh \inst{2} \and J. Alves \inst{3} \and E. Guenther \inst{4}}

\offprints{Markus Mugrauer, markus@astro.uni-jena.de}

\institute{Astrophysikalisches Institut, Universit\"at Jena, Schillerg\"a{\ss}chen 2-3, 07745 Jena, Germany \\
\and Tel Aviv University, Tel Aviv 69978, Israel \\
\and European Southern Observatory, Karl-Schwarzschild-Str. 2, 85748 Garching, Germany \\
\and Th\"uringer Landessternwarte Tautenburg, Sternwarte 5, 07778 Tautenburg, Germany \\}

\date{Received ??? ; Accepted ???}

\abstract{We report on the detection of a new low-mass stellar companion of HD\,75289, a G0V star
that harbors one known radial-velocity planet (Udry et al. 2000). Comparing an image of 2MASS with
an image we obtained with SofI at the ESO 3.58\,m NTT three years later, we detected a co-moving
companion located 21.465$\pm$0.023\,arcsecs (621$\pm$10\,AU at 29\,pc) east of HD\,75289. A second
SofI image taken 10 months later confirmed the common proper motion of HD\,75289\,B with its host
star. The infrared spectrum and colors of the companion are consistent with an M2 to M5
main-sequence star at the distance of HD\,75289. No further (sub)stellar companion down to
H\,=\,19\,mag could be detected. With the SofI detection limit we can rule out additional stellar
companions beyond 140\,AU and substellar companions with masses m\,$\ge$\,0.050\,$M_{\sun}$ from
400\,AU up to 2000\,AU.

}

\maketitle

\section{Introduction}

More than 100 extrasolar planets were discovered so far. Some of these planets were found in
multiple stellar systems. These planets are of particular interest, because they could provide some
hints about the possible implication of stellar multiplicity on planet-formation and on the
stability and evolution of planet orbits. A first indication of such possible influence could be
the apparent difference between the mass-period relation for planets in systems with only one star
and that of planets in multiple stellar systems (Zucker \& Mazeh (2002)). Furthermore Eggenberger
et al. (2004) pointed out that planets orbiting in multiple stellar systems tend to have a very low
eccentricity when their period is shorter than about 40 days.

Several groups have already searched for close (sub)stellar companions of the radial-velocity (RV)
planet host stars using adaptive optics.  Those searches leave out, however, an interesting regime
of companions, with separation up to $\sim 1000$ AU, not accessible to those searches because of
their small field of view (FOV). By using relatively wide field images and going relatively deep
(see section 2), we are able to detect wide (sub)stellar companions which could not be found so far
by the less sensitive all-sky IR surveys like 2MASS\footnote{2MASS: 2 Micron All Sky Survey} or
DENIS\footnote{DENIS: Deep Near Infrared Survey of the Southern Sky }.

Therefore, at the end of 2000, we have started an observing program to search for unknown wide
(sub)stellar companions of all stars known to harbor giant planets. So far, we have obtained a
first image for most of our target stars with the 3.8\,m UKIRT\footnote{UKIRT: United Kingdom
InfraRed Telescope} on Hawaii (northern sample) and the 3.58\,m ESO NTT\footnote{NTT: New
Technology Telescope} in Chile (southern sample). In most cases the sensitivity of the IR cameras
is sufficient to detect substellar companions with a separation down to the seeing limit
($\sim$1\,$''$). This implies that we are sensitive to companions with projected separation from
$\sim$100\,AU up to several 1000\,AU. For young RV planet host stars like $\iota$\,Hor (HD\,17051)
or $\epsilon$\,Eri (HD\,22049), with an age of only a few tens to hundred Myrs, even wide planetary
companions can be detected. The sensitivity is achieved by an observing strategy that avoids
saturation close to the host star ($\rightarrow$ detection of close companions) and by using
relatively large array IR detectors ($\rightarrow$ large FOV of more than 100\,arcsecs) for the
detection of wide companions.

Our effort already yielded one new astrometric confirmation in the northern sample. We could detect
common proper motion of a companion of the star HD\,89744 (Mugrauer et al. 2004 AN submitted),
suggested by Wilson et al. 2001. The companion is separated by about 2500\,AU from its host star,
with an effective temperature ($T_{eff}$) about 2200\,K and a mass between 0.072 and
0.081\,$M_{\sun}$, depending on the evolutionary model and the assumed age. HD\,89744\,B is either
a very low mass stellar or a heavy brown dwarf companion to a RV planet host star.

In this paper we report astrometric and spectroscopic evidence for a new stellar companion found in
our southern survey around the G0V star HD\,75289, for which Udry et al. (2000) found a planet with
m\,sin\,i\,=\,0.42\,$M_{Jup}$ in a 3.51 day orbit.

\section{Imaging, Data Reduction and Calibration}

Our own observations of HD\,75289 were obtained in the H band (1.6\,$\mu$m) with the 3.58\,m ESO
NTT. This telescope is equipped with active optics which dramatically reduced dome and telescope
seeing, yielding images with the seeing limit of the atmosphere. The IR detector is SofI
\footnote{SofI: Son of Isaac}, a 1024x1024 HgTeCd detector with 18\,$\mu$m pixel and a pixel scale
of approximately 0.144\,arcsecs in the so-called small field mode (147\,arcsecs FOV). To reduce
saturation by the bright primary we chose individual integration time to be as short as possible
(1.2\,s). To reach high sensitivity (i.e. a high limiting magnitude for the detection of faint
companions), the total integration time was around 10\,min, composed of many short integrated
images.

The auto-jitter technique of the NTT telescope was applied to delete the IR sky background from
each raw frame. The data-reduction is done with the ESO pipeline \textsl{ECLIPSE}\footnote{ECLIPSE:
ESO C Library for an Image Processing Software Environment}. All images were flat fielded with a
special dome flat image, provided by the NTT science team. At 1\,arcsec seeing the detection limit
($S/N$\,=\,3) is 19\,mag in H for 10\,min of total integration.

For calibration we identify 2MASS objects also detected on our NTT images. We use the coordinates
of those objects from the 2MASS point source catalog to determine the NTT pixel scale. We do so on
each NTT image and obtain the mean pixel scale for each NTT run. The averaged pixel scale of all
runs is 143.66$\pm$0.15\,mas (only 0.1\% relative uncertainty). With the pixel scale for each run,
we can determine the positional difference (separation) between any two stars, for the 1st, and 2nd
epoch. The separations between non-moving background stars do not change with time. Using those
non-moving background stars, we can then determine the proper motion of stars moving thru the
field. The precision of this method depends on the precision of gaussian centering per star and on
the number of the stars used. One can achieve $\sim 1/100$ of a pixel with gaussian centering and
special astrometric care (e.g. Pravdo\,\&\,Shaklan (1996)). In our study, we have achieved a
precision of $\sim$\,1/10 of a pixel ($\sim$\,20\,mas), well enough for measuring the proper
motions of our relatively nearby target stars and their co-moving companions.

\begin{figure} [htb]
\centering\resizebox{6.8cm}{!}{\includegraphics{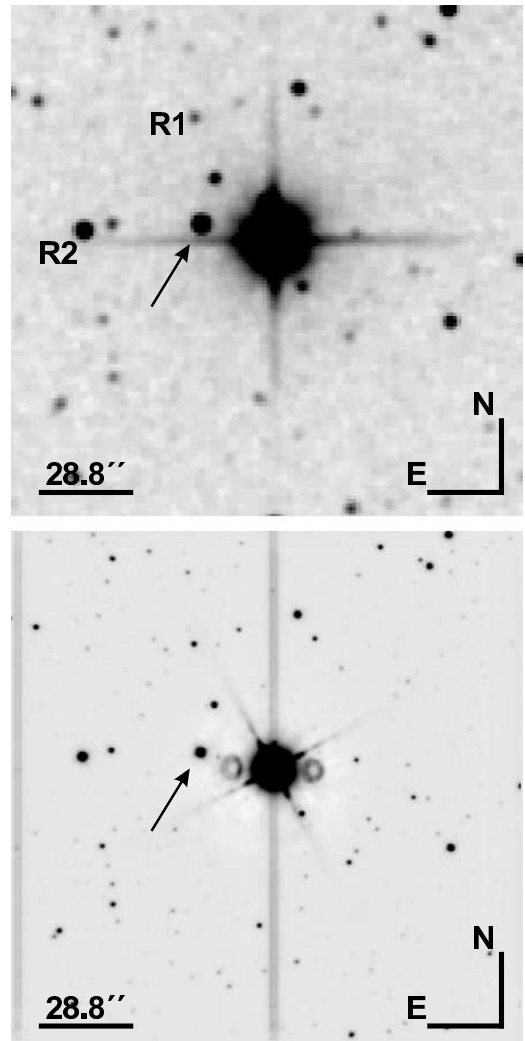}} \caption{H band images of HD\,75289
(central bright star) from 2MASS (02/99) at the top and our first epoch NTT/SofI image (01/02). The
total integration time is 10\,min. The co-moving companion is located 21.5\,arcsecs east to
HD\,75289 and is also visible in the 2MASS image (marked object). The stars R1 and R2 are used in
Fig.\,4 as comparison stars.} \label{image}
\end{figure}

\section{Astrometry}

In our search of wide (sub)stellar companions we have to examine hundreds of faint objects close to
the RV planet host stars. Most of those objects will emerge as ordinary background stars, randomly
located close to, but far behind the target stars. On the other hand, bound companions share the
proper motion of the host stars. This is so because the orbital motions of wide companions with
separations $\geq$\,100\,AU are small compared to their much faster common proper motions. An
astrometric survey will find these co-moving companions with only 2 images taken with some epoch
difference, depending on the astrometric accuracy and the proper motion of the primary stars. Hence
astrometry is a very effective tool for companion searches.

In a first step of our study of HD\,75289 we compared our first epoch NTT image with the 2MASS one.
The proper motion of the bright enough objects can be derived by comparing their position in the
2MASS and the NTT images. The 2MASS images are accurate enough for the detection of co-moving
companions, as the proper motion of HD\,75289, is large enough.

The proper motion of most stars, as derived from the 2MASS/NTT astrometry and the given epoch
difference of 2.9\,yr, were very small. Only one star had large proper motion,
$\mu_{\alpha}$\,=\,1\,$\pm$\,24\,mas and $\mu_{\delta}$\,=\,-236\,$\pm$\,22\,mas per annum,
consistent with the well known Hipparcos proper motion of HD\,75289
($\mu_{\alpha}$\,=\,-20.50\,$\pm$\,0.49\,mas/yr and $\mu_{\delta}$\,=\,-227.68$\pm$\,0.44\,mas/yr).
It is clear that this star is a co-moving companion of the RV planet host star. We therefore denote
this star as HD\,75289\,B.

However, the 2MASS limit is approximately 2.5 magnitudes brighter than the NTT limit, hence the
motion of the faint companion candidates can be investigated only with a second epoch NTT
observation (see 2MASS and SofI/NTT image in Fig.\,\ref{image}).

\begin{figure} [htb] \resizebox{\hsize}{!}{\includegraphics{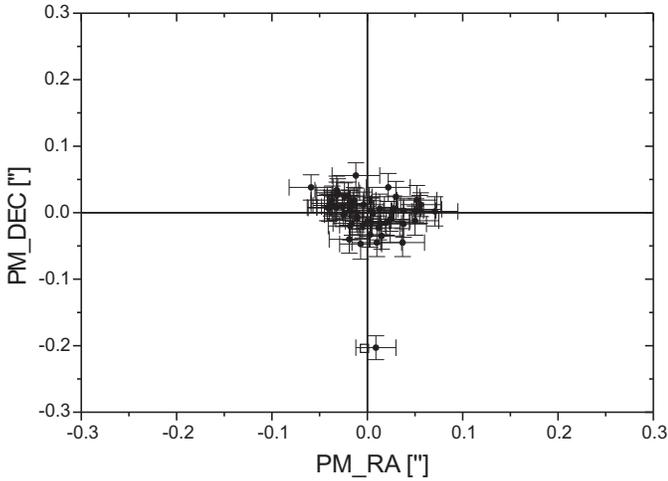}} \caption{Result of the
astrometry obtained by comparing two NTT images from epochs 01/02 and 12/02. The formal proper
motion of all detected objects around HD\,75289\,A are shown in the diagram. All of them have
negligible proper motions, similar in size as the astrometric uncertainty ($\sim$20\,mas) hence
they are very slowly moving background objects. Only HD\,75289\,B (bottom with error bars) shares
the proper motion of HD\,75289\,A (square) which is well known from the Hipparcos astrometry. The
proper motion of the bright primary star A is not measured in our NTT images because of
saturation.} \label{astro}
\end{figure}

Due to the large number of stars in the NTT FOV (see Fig.\,\ref{image}) several non-moving
background stars are detected and the proper motion can be determined with a precision in the order
of $\sim$20\,mas (see Fig.\,\ref{astro}). Due to PSF saturation the proper motion of HD\,75289\,A
cannot be measured accurately in both NTT images, but can be calculated for the given epoch
difference with Hipparcos data of the stellar parallax, yearly proper motion and equatorial
coordinates (square in Fig.\,\ref{astro}).

\begin{figure} [htb] \resizebox{\hsize}{!}{\includegraphics{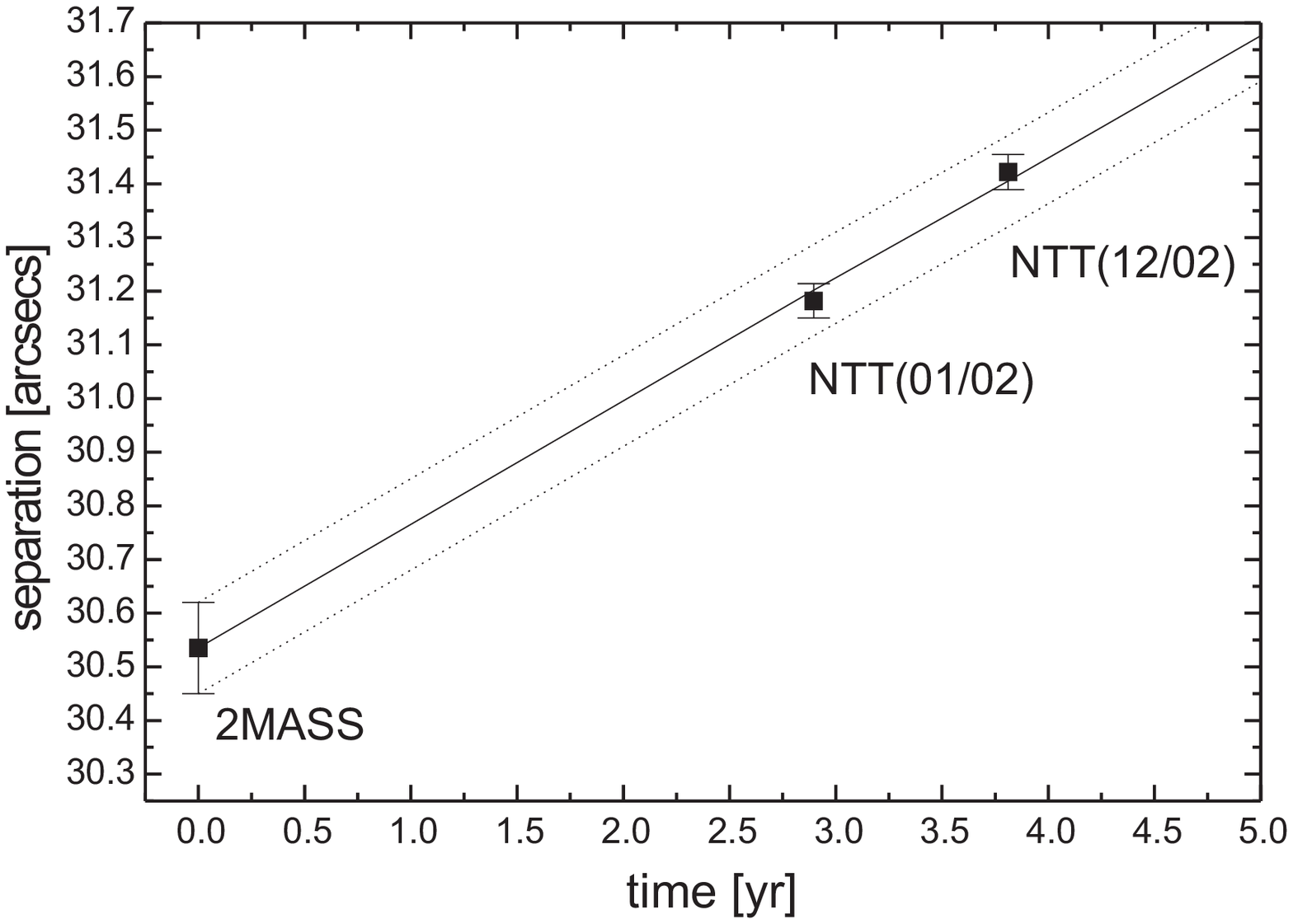}}
\newline\resizebox{\hsize}{!}{\includegraphics{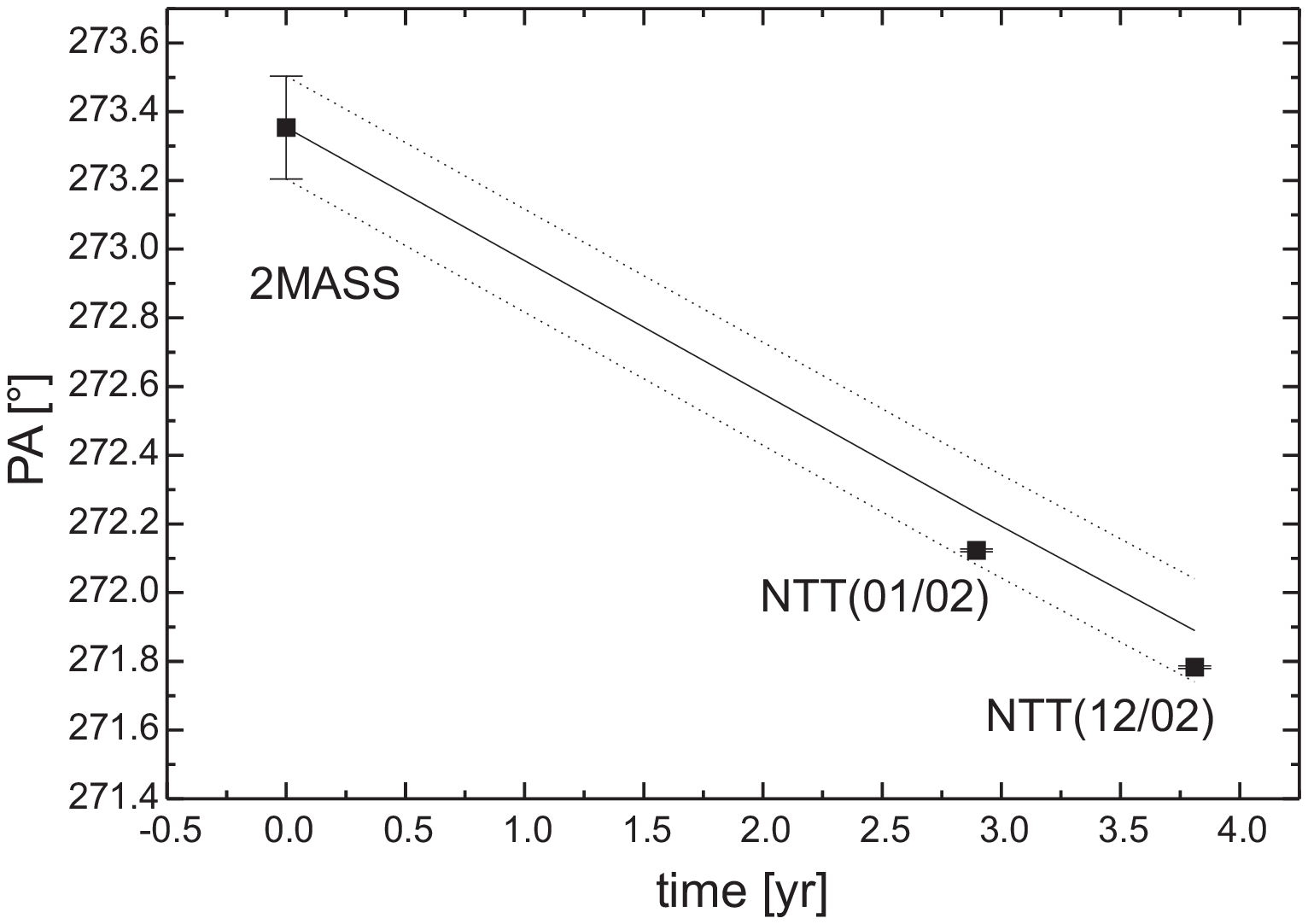}}\caption{Proper motion of HD\,75289\,B
for all three epochs. We measure the distance between R1 and HD\,75289\,B (top) and the position
angle PA of HD\,75289\,B measured from R2. See Fig.\ref{image} for the stars R1 \& R2. Due to the
motion of HD\,75289\,B relative to the reference stars both values are changing following the
predicted curves (straight lines) for a co-moving companion to HD\,752898\,A. The astrometric
uncertainty is illustrated with dotted lines.} \label{astro1}
\end{figure}

In addition we illustrated the proper motion of HD\,75289\,B over all three epochs with two
reference stars R1 and R2 (see Fig.\ref{image} and Fig.\ref{astro1}).

\section{Photometry}

Table\,\ref{photo} gives the apparent magnitude of HD\,75289\,A\,\&\,B in JHK$_{S}$ derived by
2MASS, together with our derivation of H$_{Sofi}$=11.224$\pm$0.040. Our result is consistent with
the 2MASS photometry. From the known spectral type of HD\,75289\,A and hence, its expected
intrinsic B-V color (B-V=0.58 from Kenyon~\&~Hartmann (1995)) and its published B-V color (from
Hipparcos B-V\,=\,0.578$\pm$0.003\,mag), we find that interstellar absorption is negligible, as
expected for nearby stars. With J-K from 2MASS and color to temperature conversion from
Kenyon~\&~Hartmann (1995) we can derive a $T_{eff}$ between 5800 to 6380\,K
(J-K\,=\,0.355$\pm$0.029) for the primary, which is consistent with its published spectral type G0V
(Udry et al. 2000). We obtained 3210 to 3860\,K (J-K\,=\,0.907$\pm$0.047) for the companion, hence
spectral type between M0 and M5. We used the 2MASS color transformations of Carpenter (2001) to
convert J-K$_{S}$ from 2MASS to J-K of Bessel~\&~Brett which is similar to Johnson.

\begin{table} [htb]
\caption{Photometry for HD\,75289\,A\,\&\,B. The 2MASS Point Source catalog yield apparent
JHK$_{S}$ magnitudes which are confirmed in H with our SofI/NTT images.}
\begin{center}
\begin{tabular}{l|c|c}
band& $m_{A}$ & $m_{B}$\\
\hline
J & 5.346$\pm$0.019 & 11.750$\pm$0.036\\
H & 5.187$\pm$0.031 & 11.181$\pm$0.031\\
H$_{Sofi}$ & - & 11.224$\pm$0.040\\
K$_{S}$ & 5.012$\pm$0.020 & 10.879$\pm$0.027\\
\end{tabular}
\label{photo}
\end{center}
\end{table}

\section{Spectroscopy}

To confirm the spectral type of the companion we obtained IR spectra of HD\,75289\,A\,\&\,B in June
2003 with SofI in spectroscopic mode. We used long slit spectroscopy with a slit-width of one
arcsec and the red grism covering the wavelength range from 1.53 to 2.52\,$\mu$m. The dispersion
was 10.22\,\AA~per pixel with a IR HgCdTe detector in the large field mode (288\,mas pixel scale).
The resolving power is $\lambda/\Delta \lambda \approx 588$.

Background subtraction was obtained by nodding between two positions along the slit, as well as by
a small jitter around those two positions, to avoid that individual pixels see always the same part
of the sky. Eighteen individual spectra, each with an integration time of 30\,s, were averaged,
i.e. a total integration time of 9\,min. All images were flat fielded with a standard dome flat and
wavelength calibrated with a Xe lamp. We used standard IRAF routines for background subtraction,
flat fielding and averaging all individual spectra.

\begin{figure} [htb]
\resizebox{\hsize}{!}{\includegraphics{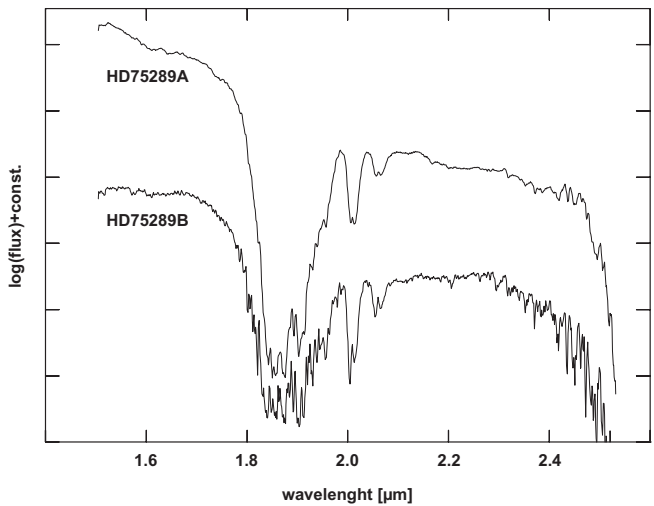}} \caption{Relative flux of HD\,75289\,A\,\&\,B.
The drop at 1.85\,$\mu$m is due to strong water absorption as well as some telluric lines. From a
black body fit, we derive a $T_{eff}$ of the companion between 3250K and 3500K, coincident with a
spectral type M2 to M5.}\label{s1}
\end{figure}

The companion and the primary star were both located on the slit, and spectra of both objects were
taken simultaneously. $T_{eff}$ of HD\,75289\,A is well known, hence a black body function with the
given $T_{eff}$ (6030K) can be used to determine the response function of the spectrograph, which
is needed to obtain relative flux calibrated spectra of both objects. In Fig.\ref{s1} we show the
relative flux calibrated spectra of HD\,75289\,A\,\&\,B. The continuum of the companion is much
flatter than the primary continuum, consistent with a cooler photosphere. From a black body fit on
the continuum of the HD\,75289\,B we determine its $T_{eff}$ to be in the range between 3250\,K and
3500\,K, hence spectral type M2 to M5.

\begin{figure}[htb]\resizebox{\hsize}{!}{\includegraphics{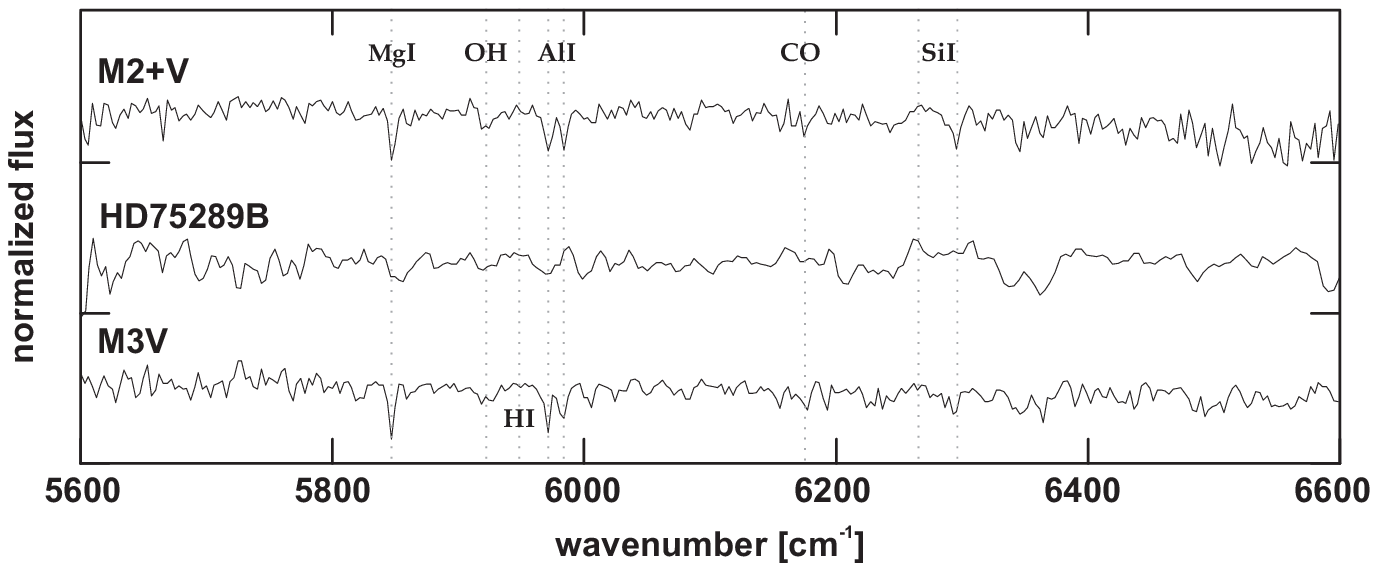}}
\resizebox{\hsize}{!}{\includegraphics{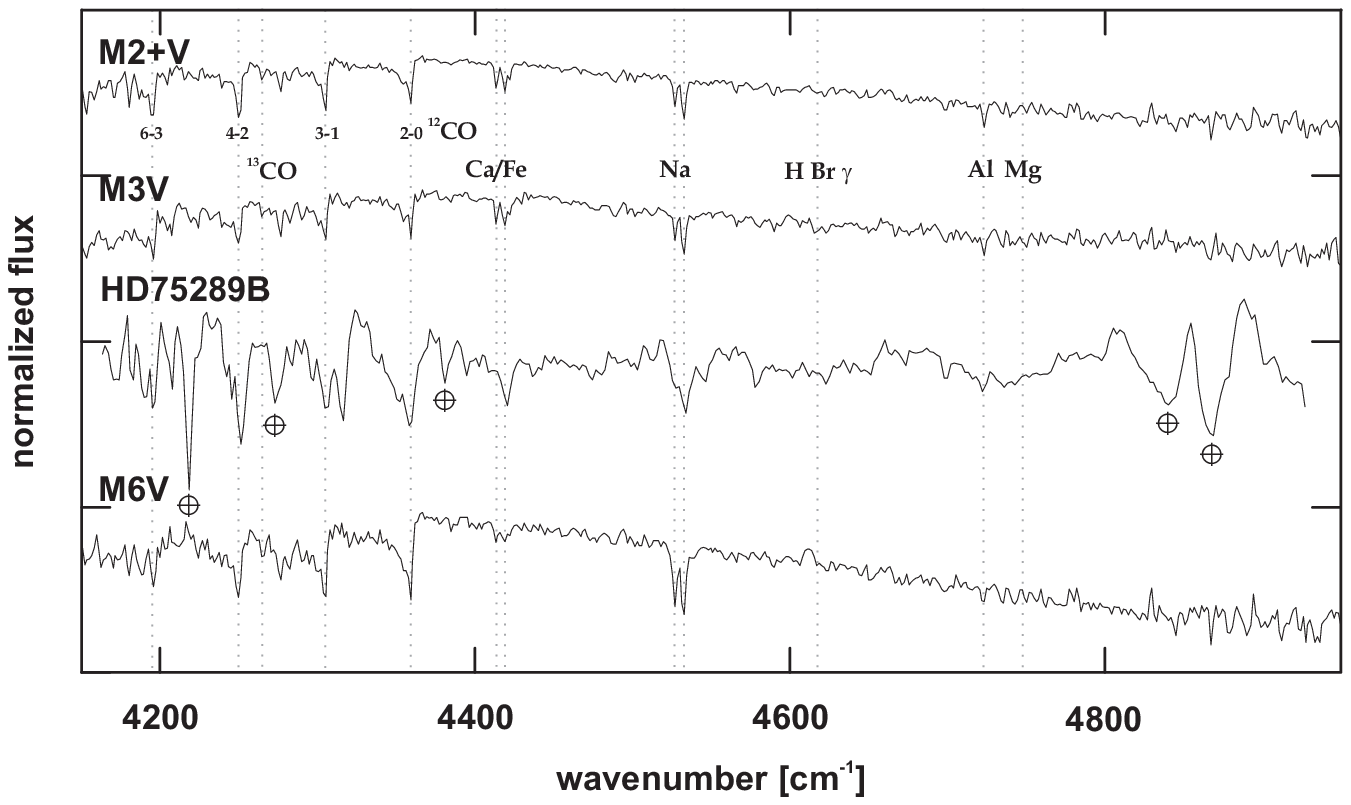}} \caption{Normalized H and K band spectra of
HD75289\,B, compared with spectra of GJ\,411 (M2+V), GJ\,725 (M3V) and Wolf\,359 (M6V) from Meyer
et al.(1997/98).} \label{spec}
\end{figure}

Fig.\,\ref{spec} shows the normalized H and K band spectra of HD\,75289\,B. The most striking
luminosity-sensitive feature in the H band is the second-overtone CO band head $[\nu,\nu^{'}]$ at
6177\,cm$^{-1}$, which is found in the spectra of K and M stars. The spectra of HD\,75289\,B shows
CO molecular lines which are fainter than Mg (5844\,cm$^{-1}$), typical for red dwarfs and
coincident with the JHK absolute magnitudes derived in section\,4. Hydrogen line at
5950\,cm$^{-1}$, which can be found only in stars earlier than K3, is not visible in the companion
spectra and the Al feature at 5973\,cm$^{-1}$ is strong as the Mg line. Both comparisons serve as
evidence for a spectral type cooler than M1V. The Si line at 6264\,cm$^{-1}$ is not apparent and Si
at 6292\,cm$^{-1}$ is faint comparable with the OH ($\Delta\nu=2$) molecular features at
5920\,cm$^{-1}$, typical for a spectral type M3V.

The strongest apparent lines in the companion spectrum in K are from molecular bands of the first
CO overtone extending from 4360\,cm$^{-1}$ to the low frequency side of the spectrum. In addition,
$^{13}$CO at 4260\,cm$^{-1}$ and H Br\,$\gamma$ at 6297\,cm$^{-1}$ are not apparent, all argue for
a dwarf cooler than K2V. The CO bands are a bit stronger than the Ca/Fe (4415\,cm$^{-1}$) as well
as Na (4530\,cm$^{-1}$) atomic features and Ca/Fe is weaker than Na. The Al line at 6720\,cm$^{-1}$
is faint but the Mg line at 4750\,cm$^{-1}$ is missing, typical for spectral types cooler than M3V.

The detected features in the spectrum of HD\,75289\,B in H and K, the black body fit of the
continuum, and the companion JHK colors, are all consistent with a spectral type of M2V to M5V,
i.e. $T_{eff}$ is in the range of 3240 to 3580\,K.

\section{Discussion}

HD\,75289 is a bright G0 dwarf (V=6.36\,mag) located at a distance of 28.94$\pm$0.47\,pc (distance
module 2.308$\pm$0.036\,mag). Its apparent JHK colors are typical for a G0V star at the given
distance. Thus, the super-giant classification given in Simbad is invalidated, as pointed out by
Udry et al. (2000). The same group discovered an extrasolar planet with a minimum mass of 0.42
M$_{Jup}$ which revolves around its parent star in a nearly circular orbit (e=0.054, a=0.046\,AU).

\begin{figure} [htb] \resizebox{\hsize}{!}{\includegraphics{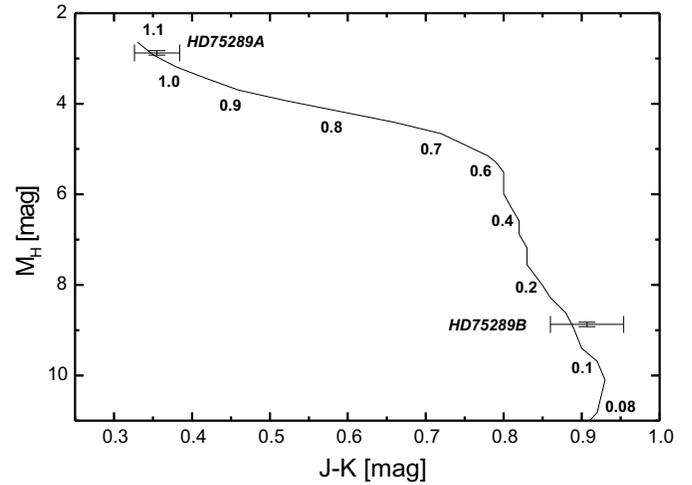}} \caption{The
color-magnitude diagram with the isochrone for 5\,Gyrs from Baraffe et. al (1998) with [M/H]=0,
mixing length parameter $\alpha$=1 and He abundance Y=0.25. The primary and its companion are
included in the diagram with their uncertainties in magnitude and color. Masses are indicated as
numbers in solar masses.} \label{hrd}\end{figure}

HD\,75289\,B is clearly co-moving with HD\,75289\,A and the color-magnitude relation agrees with
the assumption that both objects are at the same distance (Fig.\,\ref{hrd}). With Baraffe et al.
(1998) models, the JHK colors from Sect.\,3 and the given distance module, we can derive the mass
of HD\,75289\,B to be 0.135$\pm$0.003\,M$_{\odot}$, see Fig.\,\ref{hrd}. The system age is roughly
5\,Gyrs (Udry et al. 2000). Note that the age uncertainty of the primary does not play an important
role in the derivation of the mass of HD\,75289\,B, because the IR magnitudes for such low-mass
stellar objects decrease very slowly from 1 to 10\,Gyrs. The given uncertainty of the companion
mass is derived only form the magnitude errors. Inaccuracies of the used theoretical model were not
considered here.

With the derived companion mass (0.135\,$M_{\sun}$), the primary mass ($\sim$1$M_{\sun}$) and the
companion separation 621\,AU (21.465$\pm$0.023\,arcsecs) we can compute the expected RV variation
of the primary induced by the presence of the wide companion v$\sim$\,150\,m/s with an orbital
period of $\sim$\,15000\,years. Although this is a large effect, the maximal yearly variation of
the RV is only $\sim$0.07\,m/s, too small to be detected in the foreseen future.

\begin{figure}[htb] \resizebox{\hsize}{!}{\includegraphics{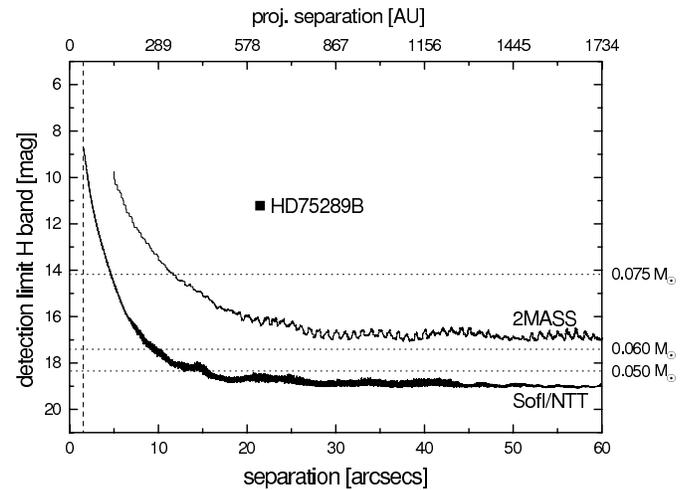}} \caption{The limiting
H magnitude versus separation from HD\,75289\,A for our NTT image shown in Fig.\ref{image} and
2MASS. The corresponding projected separation in AU is shown on the upper x-axis. Saturation occurs
within 1.5\,arcsecs (43\,AU) (see vertical dashed line) hence a companion search is impossible
there. The detection of all stellar companions is feasible beyond 4.7\,arcsecs (136\,AU). The right
y-scale shows the predicted absolute H magnitudes for substellar objects from Baraffe et al. (2003)
models for an age of 5\,Gyrs. The 3$\sigma$ detection limit is 19\,mag in H, hence substellar
companions down to 0.050\,$M_{\sun}$ can be found beyond 15\,arcsecs (434\,AU).} \label{dynamic}
\end{figure}

Fig.\ref{dynamic} shows the NTT detection limit which is 19\,mag in H and enables the detection of
substellar companions down to M$_{H}$=16.7\,mag around HD\,75289\,A (m$\ge$44\,M$_{Jup}$ according
to Baraffe et al. 2003). Objects up to 68\,arcsecs were observed twice but no further co-moving
companion could be identified. Further stellar companions (m$\ge$75\,M$_{Jup}$) can be ruled out
for a projected separation from 136\,AU up to 1968\,AU.

\acknowledgements {We would like to thank the technical staff of the ESO NTT for all their help and
assistance in carrying out the observations. Furthermore, we would like to thank M. Fern\'andez,
A.~Seifahrt, A.~Szameit and C.~Broeg who have carried out some of the observations of this project.
We made use of the 2MASS public data releases as well as the Simbad database operated at the
Observatoire Strasbourg. This work was partly supported by the Israel Science Foundation (grant no.
233/03)}

{}

\end{document}